# Chronic Obstructive Pulmonary Disease Prediction Using Deep Convolutional Network

Shahran Rahman Alve[1*], Muhammad Zawad Mahmud[1], Samiha Islam[1] and Mohammad Monirujjaman Khan[1]

[1]Department of Electrical and Computer Engineering, North South University, Bashundhara R/A, Dhaka 1229, Bangladesh
[*]Corresponding Author: Shahran Rahman Alve. Email: shahran.alve @northsouth.edu

**Abstract:** Artificial intelligence and deep learning are increasingly applied in the clinical domain, particularly for early and accurate disease detection using medical imaging and sound. Due to limited trained personnel, there is a growing demand for automated tools to support clinicians in managing rising patient loads. Respiratory diseases such as cancer and diabetes remain major global health concerns requiring timely diagnosis and intervention. Auscultation of lung sounds, combined with chest X-rays, is an established diagnostic method for respiratory illness. This study presents a Deep Convolutional Neural Network (CNN)-based approach for the analysis of respiratory sound data to detect Chronic Obstructive Pulmonary Disease (COPD). Acoustic features extracted with the Librosa library, including Mel-Frequency Cepstral Coefficients (MFCCs), Mel-Spectrogram, Chroma, Chroma (Constant Q), and Chroma CENS, were used in training. The system also classifies disease severity as mild, moderate, or severe. Evaluation on the ICBHI database achieved 96% accuracy using 10-fold cross-validation and 90% accuracy without cross-validation. The proposed network outperforms existing methods, demonstrating potential as a practical tool for clinical deployment.

**Keywords:** COPD; Sound; Database; Convolutional Neural Network; Deep Learning; Detection; CNN; Accuracy; Prediction.

## 1 Introduction

The clinical benefits dimension is critical and one of the most effective ways to distinguish healthcare projects. It remains a central area of investment, aiming to improve diagnosis, therapy, and overall quality of care. Clinical equipment varies in purpose depending on its sensitivity, clarity, and design, and is often reviewed by experts or translators before being made available to broader use [1].

Historically, interpretation of the large volumes of clinical data produced by automated systems relied primarily on human expertise. Chronic Obstructive Pulmonary Disease (COPD) comprises a group of lung disorders that obstruct airflow through narrowed airways, impairing oxygen intake and carbon dioxide release. The two principal diseases underlying COPD are emphysema and chronic bronchitis [2]. Patients commonly exhibit both, with rapid deterioration possible. For example, chronic bronchitis requires management of breathing with markedly reduced lung capacity [3].

COPD arises mainly from smoking, genetic predisposition [4], environmental pollution, and other risk factors. It remains difficult to detect early or with full accuracy. Standard diagnostic methods include [5]:

- **Respiratory muscle tests**: Assess pulmonary function and oxygen exchange. Spirometry is the most widely used, measuring airflow during exhalation.
- **Chest X-ray**: Identifies respiratory pathologies, with emphysema being a leading cause of COPD.
- **CT scan**: Less frequently applied but critical when other methods fail to confirm diagnosis. CT imaging aids in treatment assessment.
- **Arterial blood gas test**: Evaluates oxygen saturation in arterial blood during respiration.

**Limitations of current methods**:

- Spirometry is contraindicated in patients with severe cardiac disease or recent cardiac surgery.
- Testing may induce shortness of breath, nausea, or dizziness.
- X-rays and CT scans expose patients to ionizing radiation.

Because of these limitations, research has examined the use of artificial intelligence and machine learning to enhance diagnostic accuracy for COPD. Algorithms have been applied to medical images and respiratory sounds to support clinical decision-making. The related work highlights systems designed to automate lung disease detection. For example, the Breath Monitoring System [6] introduced a rapid inspection device for COPD detection. Another study used an artificial neural network (ANN) with a backpropagation multilayer perceptron to classify respiratory sound events in asthma and COPD patients [7]. The nonlinear nature of ANN yielded better results than conventional methods. Reported precision and recall were 77.1% and 78.0% for key events, and 83.9% and 83.2% for other events, with an overall average of 81.0%. Improvements are expected through refined classifiers and deep learning techniques. Simplified frameworks for chronic illness management have also been proposed to enable continuous monitoring of COPD patients [8]. One system allowed physicians to track patient conditions in real time. Hybrid classifiers combining support vector machines, random forests, and rule-based methods achieved early COPD detection accuracy of up to 94%. In another study, a computer-based approach was developed to analyze stethoscope-recorded respiratory sounds [9]. The device collected three types of breathing sounds from 60 patients. A deep learning model comprising six convolutional layers, three max-pooling layers, and three fully connected layers was trained using log-scaled Mel spectrograms, divided into 23 frames. The model was tested on new data from 12 patients and evaluated by five respiratory specialists. Applications include telemedicine and self-screening. Aykanat et al. (2017) introduced a low-cost electronic stethoscope capable of recording lung sounds in clinical and non-clinical settings. Using this device, 17,930 lung sound samples were collected from 1,630 participants [10]. The study applied two machine learning methods: Mel-frequency cepstral coefficient (MFCC) features in a Convolutional Neural Network (CNN), and Support Vector Machine (SVM) models trained on spectrogram images. MFCC features are widely used for audio classification, and were employed to benchmark CNN performance. Both CNN and SVM generated four outputs for respiratory sound classification: detection of rales, rhonchi, and normal breath sounds; respiratory sound pattern lists; and sound-type assignments per sample. Reported accuracies were 86% (CNN), 86% (SVM), 76% (CNN), 75% (SVM), 80% (CNN), 80% (SVM), and 62% (CNN), 62% (SVM). These results, evaluated independently, showed that CNN and SVM both achieved strong performance in spectrogram-based classification. With larger datasets, these approaches could be extended for COPD pre-screening. The same study also assessed Adaptive Neuro-Fuzzy Inference Systems (ANFIS), Multiple ANFIS (MANFIS), and Coactive ANFIS (CANFIS) for spirometry interpretation [11]. ANFIS yielded higher accuracy than conventional neural networks by combining neural learning with fuzzy inference. CANFIS achieved the best performance, with 97.5% accuracy, surpassing both ANFIS and MANFIS. Chamberlain et al. (2016) [12] investigated wheezes and crackles, the two most common abnormal lung sounds. Their models achieved ROC AUCs of 0.86 for wheezes and 0.74 for crackles. In a separate experiment, a dataset of 155 samples (55 COPD, 100 normal) was used to evaluate multilayer neural networks. A two-hidden-layer MLNN achieved 95.33% accuracy, outperforming a single-layer model. A feed-forward neural network with a log-sigmoid activation function was also tested, using supervised learning and iterative error analysis for refinement. Radial Basis Function (RBF) networks were highlighted as effective for nonlinear data, since they compute hidden layer outputs in a single step before linear optimization at the output layer. Machine learning methods show strong potential for early detection of COPD exacerbations [13]. Exacerbations are a major cause of reduced quality of life and hospitalizations. Models can identify signs of exacerbation up to 4.4 days before symptom onset. Decision tree and random forest classifiers have been validated on recorded datasets, while deep learning methods such as CNNs and LSTMs are gaining traction [14]. With sufficiently large datasets of respiratory sounds and spectrograms, robust architectures can be trained without hand-crafted features. Automated sound-based detection offers clear advantages: reproducibility, sensitivity, adaptability, and the ability to tune for different clinical priorities, such as high sensitivity in screening applications. Traditional COPD detection relies on pulmonary function tests or imaging, which are costly, time-intensive, and require expert handling. Automated image-based methods further depend on very large, high-quality datasets, which are not always available. In contrast, respiratory sound analysis provides a low-cost, non-invasive alternative that can be integrated into clinical workflows. Future extensions may link such systems with smartphones or wearable devices for continuous monitoring and early COPD detection.

This work proposes a novel CNN architecture designed for efficient and accurate COPD detection. The approach reduces computational requirements while maintaining diagnostic accuracy. Section 2 details the methodology and materials, Section 3 presents results, and Section 4 provides the conclusions.

## 2 Methods and Materials

The data were obtained from publicly available online sources. After partitioning into training and test sets, images and annotations were imported and extracted from the raw datasets. Preprocessing and enhancement methods were then applied. The following section outlines hyperparameter selection, regularization, and optimization strategies. System training and performance estimates are subsequently presented. The CNN model, implemented in Google Colab, is used to classify normal and abnormal sounds.

*2.1 Dataset Description*

This study used the Respiratory Sound Database dataset [15], a publicly available resource developed by research groups in Portugal and Greece. The dataset comprises 920 annotated recordings ranging from 10 to 90 seconds, collected from 126 patients. In total, it includes 5.5 hours of data encompassing 6,898 respiratory cycles, of which 1,864 contain crackles, 886 contain wheezes, and 506 contain both. The recordings include both clean and noisy samples to simulate real-world conditions. Patients represent all age groups, including children, adults, and the elderly.

*2.1.1 Data Preprocessing*

The dataset contained inconsistencies and required preprocessing. Using the Python library **Librosa**, all sound recordings were standardized to a duration of 20 seconds. Feature extraction was based on five characteristics.

**Mel-Frequency Cepstral Coefficients (MFCCs)** represent the power spectrum of a signal by applying a cosine transform to a log-scaled power spectrogram along a non-linear mel scale. They capture phonetic information by reflecting vocal tract behavior, which complicates their interpretation for respiratory sound analysis.

To generate the **Mel-Spectrogram**, pressure signals over time were transformed into frequency space using the Fast Fourier Transform, mapped onto the mel scale, and represented as amplitude over time. These highlights temporal changes in spectral power.

**Chroma features** (pitch class profiles) track pitch content and are widely used in music analysis. They are relevant for respiratory studies since breathing sounds exhibit distinct pitch variations.

**CENS features** (Chroma Energy Normalized Statistics) are sensitive to dynamics, tone, and articulation, making them suitable for sound classification and retrieval tasks. For uniformity, the number of coefficients was fixed at *n = 40* across all feature sets.

*2.1.2 Data Augmentation*

Since the number of COPD and non-COPD tests was often similar, data augmentation techniques were applied to increase the number of non-COPD samples. A convolutional neural network (CNN) was implemented using Keras and TensorFlow. The architecture consists of an input data layer, multiple Convolution2D layers, Dropout layers, MaxPooling2D layers, and a Dense layer. In convolutional layers, filters are the key components. Each filter slides across the input and produces a feature map by detecting local patterns through convolution. Each convolutional layer is followed by a MaxPooling2D layer, while the final stage employs a GlobalAveragePooling2D layer. Pooling reduces dimensionality by retaining critical values and decreasing the number of trainable parameters, thereby shortening training time and reducing overfitting risk. MaxPooling selects the maximum value within each region, while Global Average Pooling computes the average, and the resulting outputs are forwarded to the dense classification layer. The final layer applies Softmax activation, which normalizes the outputs into probability distributions, assigning the highest value to the most likely class.

2.2 Block Diagram

Figure 1 depicts the research method as a system diagram.

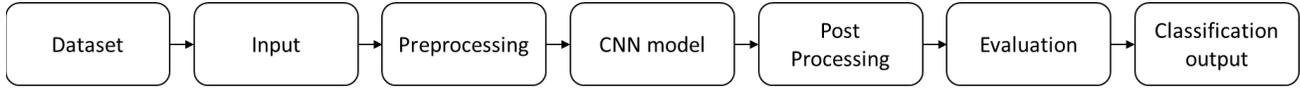

**Figure 1:** System Block Diagram

The input data has the shape (40, 862, 1), where 40 denotes the number of MFCCs, 862 the number of frames (including padding), and 1 the mono audio channel. The model begins with a Convolution2D layer (16 filters, kernel size: 2, ReLU), followed by a MaxPooling2D layer and a 20% dropout to mitigate overfitting. This is followed by a Convolution2D layer (32 filters, kernel size: 2, ReLU), a MaxPooling2D layer (pool size: 2), and another Convolution2D layer (64 filters, kernel size: 2, ReLU), again followed by MaxPooling2D (pool size: 2) and dropout (20%). The sequence continues with a Convolution2D layer (32 filters, kernel size: 2, ReLU), a MaxPooling2D layer (pool size: 2), and dropout (20%). ReLU is consistently used as the activation function to generate feature maps. Finally, the output is passed through a GlobalAveragePooling2D layer (128 units) to reduce dimensionality, followed by a dense layer for binary classification into COPD and non-COPD categories. The architecture is illustrated in Figure 2.

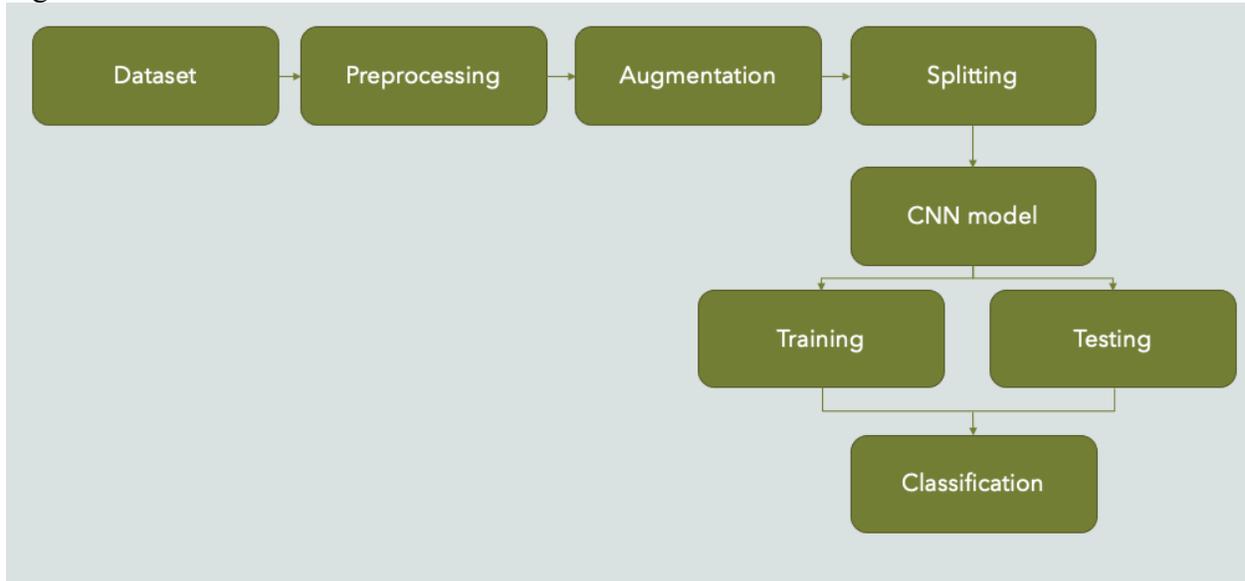

**Figure 2:** System Architecture of the Proposed Method

When compared to the image inputs, the audio data sources exhibit higher levels of noise. To address this, we employed the Adam optimizer, which is computationally efficient, memory-friendly, and well-suited for irregular data. Adam combines stochastic gradient descent with the RMSprop algorithm, enabling more effective parameter updates and facilitating robust hyperparameter tuning [16]. Compared to other optimizers such as RMSprop, SGD, Adagrad, and NAdam, Adam achieves faster convergence with more stable performance.

*2.3 Proposed System*

Convolutional layers, ReLU, and Dropout were employed in the convolution block. The convolutional layer operates similarly to max-pooling. A SoftMax function is placed in the final layer. A 2D convolutional layer applies $K$ filters (kernels) of size $(M \times N)$ to the input image, computing the dot product between kernel weights and input. The filter movement across the image is defined by the stride, while padding $(P)$ may be added to preserve border information. Dropout $(D)$ is used to mitigate overfitting. Kernels detect features, ranging from low-level patterns such as edges and lines in earlier layers to more complex features in deeper layers.

The network parameters were: filters = 64, 128, 128; kernel sizes = 3×3, 3×3; stride = [1, 1, 1]; padding = same; dropout rate = 0.2. A dropout rate of 20% was found to be optimal. Each convolutional layer is followed by a ReLU activation function, which accelerates training compared to saturating functions. ReLU outputs the input if positive and zero otherwise.

Max-pooling is used for downsampling, providing spatial invariance. The pooling layer partitions the image into non-overlapping 3×3 regions with stride 3×3, retaining only the maximum value within each region. This reduces both the number of parameters and computational cost. The proposed CNN architecture is illustrated in Figure 3.

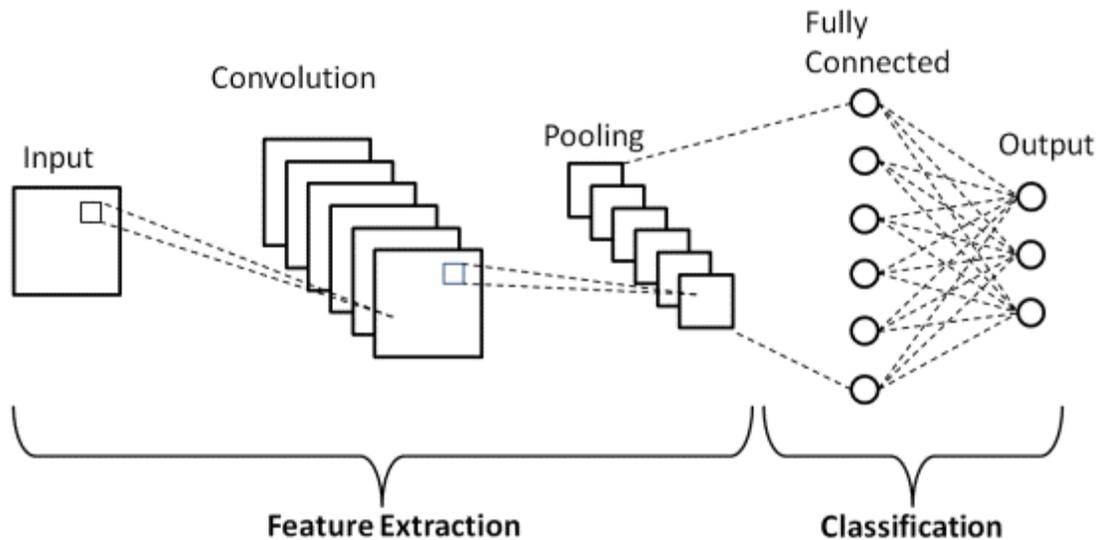

**Figure 3:** Proposed CNN Architecture

The network architecture is illustrated in Figure 3 and consists of an input, three convolutional blocks, a classification unit, and an output.

**Block 1.** The first convolution layer generates an output twice the size of the input using 3×3×64 filters with stride = 1 and padding = 1. The convolutional response is passed through a ReLU activation, followed by a 20% dropout layer. A max-pooling operation then reduces the spatial resolution by half, producing an output twice as large as the input dimension.

**Block 2.** The output of Block 1 serves as the input to Block 2. This block applies a convolution with 3×3×128 filters, stride = 1, and padding = 1. Batch normalization is used instead of dropout, enabling stable training. The convolutional response passes through a ReLU activation and a max-pooling layer, again producing an output twice the size of the input.

**Block 3.** This block mirrors Block 2, but includes a 20% dropout layer in addition to ReLU activation and max-pooling. The output, twice the size of the input, is fed into the classification unit.

**Classification unit.** The classification stage consists of four fully connected (FC) layers. The first flattens the previous output into 8192 nodes. The second and third layers are dense with 128 and 64 nodes, respectively. The final layer contains the same number of neurons as target classes and uses a SoftMax activation for classification.

*2.4 Cross-Validation*

In this study, performance was evaluated using cross-validation. To ensure robust conclusions, the dataset was divided into 90% training and 10% test images. Only the training data was used for cross-validation. As illustrated in Figure 6, the data was partitioned into ten equal subsets, with nine used for training and one for validation. Each cycle involved training on nine subsets and validating on the remaining one, ensuring that all samples contributed to validation exactly once. Test data was separated from the start to evaluate the model on previously unseen images. The cross-validation process is depicted in Figure 4.

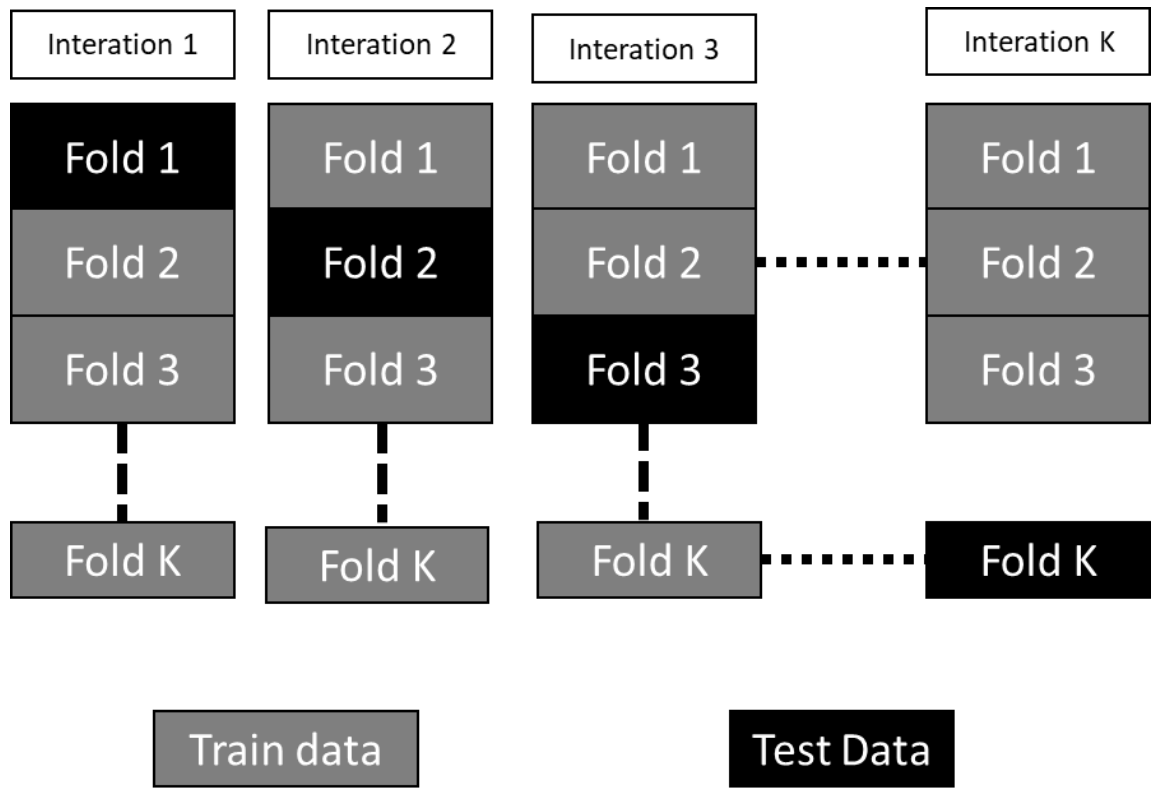

**Figure 4:** K- fold cross-validation process diagram

## 2.5 Performance Matrix

The system displayed real and expected values on a confusion matrix. The predicted outcomes of a classification model are represented by the confusion matrix. The precision, sensitivity, specificity, and accuracy numbers were calculated as follows [21], [32]:

$$Precision = \frac{TP}{TP+FP} \quad (1)$$

$$Recall = \frac{TP}{TP+FN} \quad (2)$$

$$Accuracy = \frac{TP+TN}{TP+FP+TN+FN} \quad (3)$$

$$F1 - Score = 2 * \frac{Precision * Recall}{Precision + Recall} \quad (4)$$

**True Positive (TP)** refers to the number of instances correctly predicted as positive. **True Negative (TN)** denotes the number of instances correctly predicted as negative. A **False Negative (FN)** occurs when positive instances are incorrectly predicted as negative, also known as a type II error. A **False Positive (FP)** represents the number of negative instances incorrectly predicted as positive.

## 3 Result and Analysis

We employed a bespoke CNN model in the proposed system. The dataset was used for both training and performance enhancement through 10-fold cross-validation. Without cross-validation, the model achieved a training accuracy of 98% and a validation accuracy of 95%. On evaluation, the test accuracy was 96%. With cross-validation, the test accuracy improved to 96.5%.

*3.2 Best Model Accuracy and loss*

Figures 5 and 6 show the training and validation accuracy graphs, as well as the training and validation loss graph.

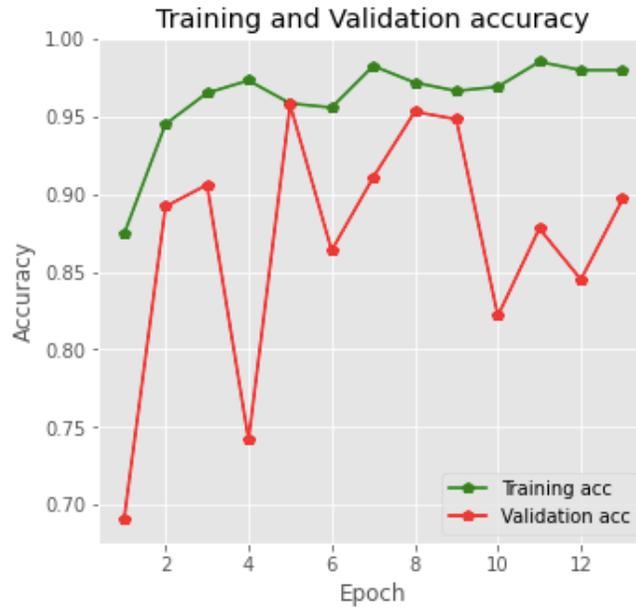

**Figure 5:** Training and Validation accuracy of the CNN model

With a learning rate of 0.001, batch size of 64, and 14 epochs, the model was trained. Performance improved across epochs, with rapid gains during the initial phases. After 5 epochs, progress slowed, and after 6 epochs, improvements were minimal.

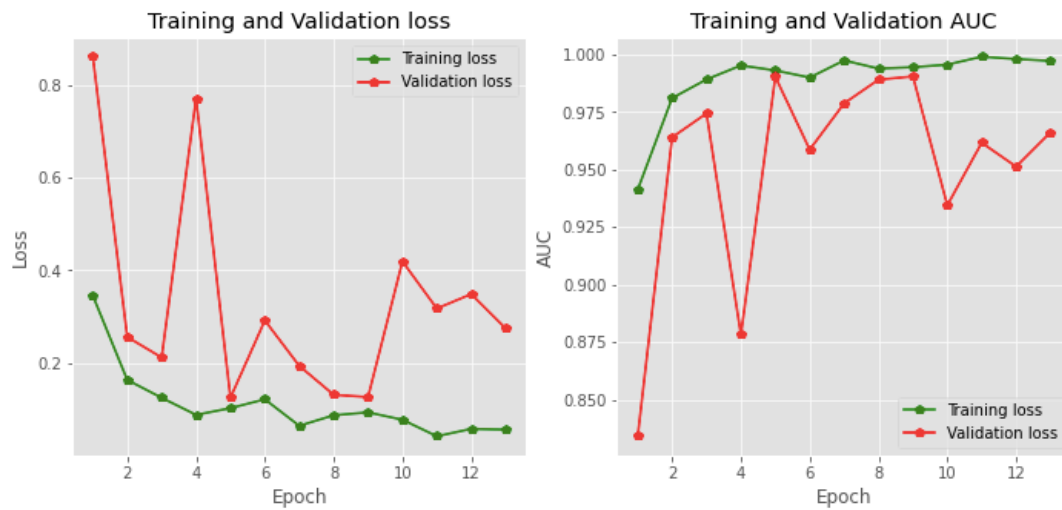

**Figure 6:** Training and Validation Loss and AUC

Figure 6 shows that the training accuracy reached 98.0 percent, while the validation accuracy was 90.0 percent. As illustrated in Figure 6, the validation loss is also greater than the training loss.

These results indicate that the model is not overfitted.

### 3.3 Other Overfitted and Low Accuracy Model Accuracy and loss

Figures 7 show the training and validation accuracy graphs, as well as the training and validation loss graph.

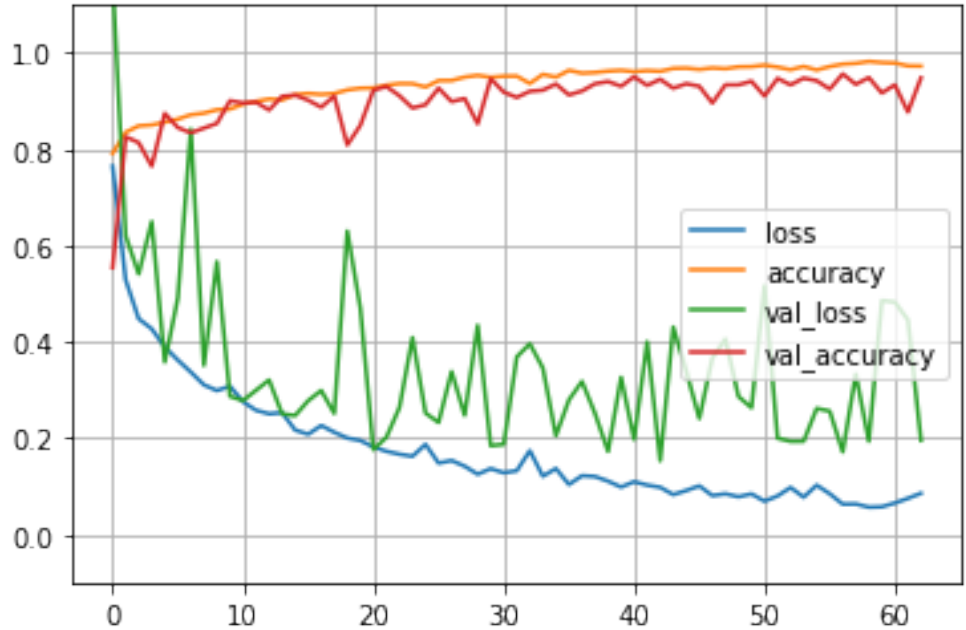

**Figure 7:** Training and Validation Accuracy and Loss of VGG 19 algorithm

With a learning rate of 0.001, 64 sample sizes, and 60 epochs, the model was trained. In this case training accuracy and validation accuracy was 89 and 88 percent respectively. The accuracy was lower than CNN model. Figure 8 shows the Inception V3 model's accuracy graph.

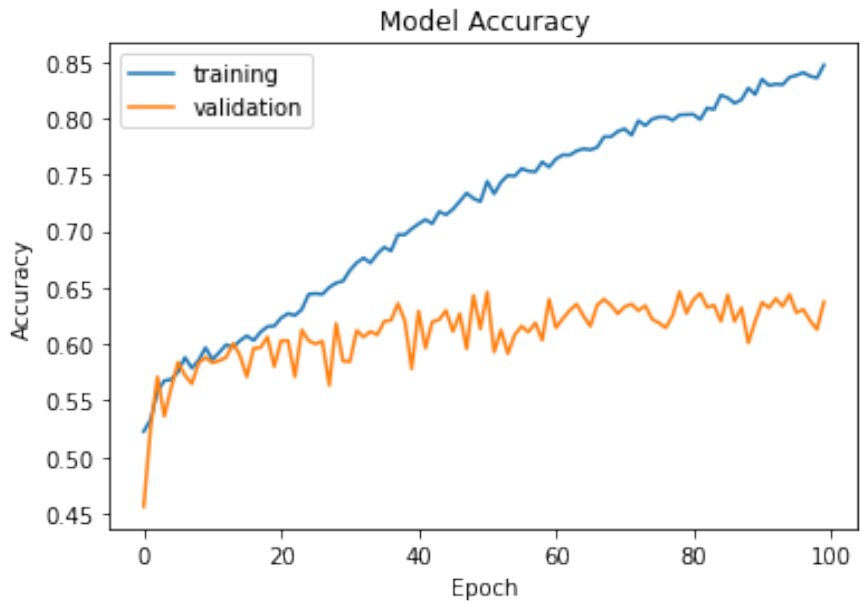

**Figure 8:** Training and Validation Accuracy of Inception V3

With a learning rate of 0.001, 64 sample sizes, and 100 epochs, the InceptionV3 was trained. In this case training accuracy and validation accuracy was 86 and 63 percent respectively. Also, with this model we received 86.90 percent test accuracy. But in this case training accuracy was much higher than validation accuracy which indicated that this model was fully overfitted.

*3.3 Audio Data Visualization*

With a 50ms FFT window size, spectrograms are produced. The spectrogram's resulting bins are multiplied by the square of their frequency. Figure 9 shows the filtering diagram.

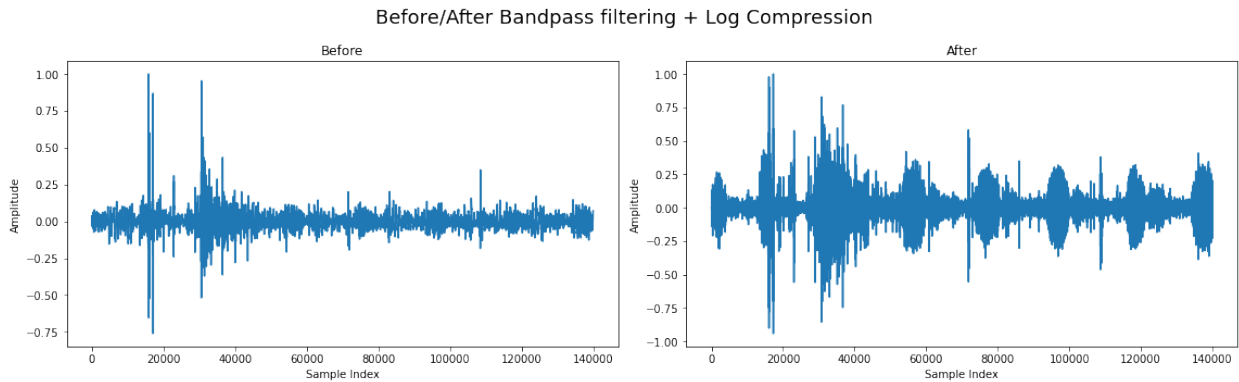

**Figure 9:** Before and After Bandpass Filtering and Log Compression

In this case, the amplitude square relation is ignored because ignoring it results in cleaner peaks. Transients are eliminated and the curve is smoothed out with a Guassian filter. On the smoothed curve, peak detection is next carried out. The timings of each peak's left and right bases are then recorded, with respect to 80% of peak height. The dataset's sound formatting image is depicted in Figure 10.

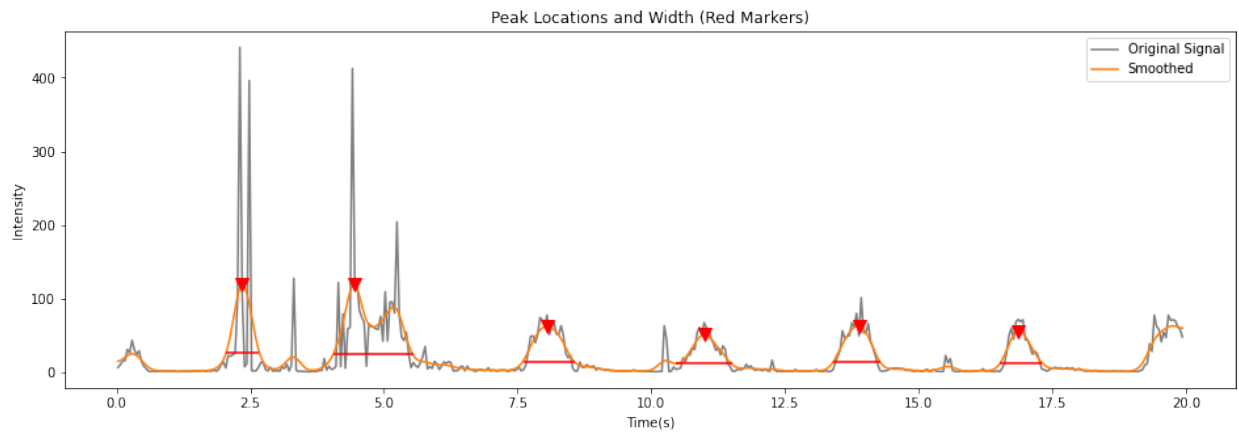

**Figure 10:** Peak and Width Location

The start and end of a breathing cycle are referenced at the left and right peak bases. A constant offset is applied to these bases to estimate cycle boundaries. Using the hand-annotated dataset, the offset values are determined by minimizing the error between predicted and annotated cycle start and end times. A multivariate optimization function is used to compute the offsets, with the objective function defined as the sum of mean errors in start and end times. Only peaks nearest to the annotated cycles are compared, as the number of detected peaks exceeds the number of annotated cycles.

*3.3 Model Classification*

Several network architectures, including VGG-19, InceptionV3, EfficientNet, and a standard CNN, were evaluated. Among these, the CNN achieved the best performance, and results are presented based on this architecture. Table 1 reports the accuracy and loss histories for all four models.

**Table 1.** Comparison of Different Deep Learning Models

| No # | Configuration | Weighted F1 score (%) | Accuracy (%) |
|---|---|---|---|
| 1 | VGG-19 | 88.58 | 88.92 |
| 2 | InceptionV3 | 86.69 | 86.90 |
| 3 | EffNet Threshold | 84.78 | 84.90 |
| 4 | **CNN** | **96.0%** | **96.0%** |

The performance of the proposed network was evaluated using augmented audio, and its generalization was assessed through 10-fold cross-validation. The model achieved an overall accuracy of 96%, outperforming existing approaches. Without cross-validation, the network reached 90% efficiency. Record-wise cross-validation yielded the highest results, with an average accuracy of 96%. Figure 7 illustrates the classification performance on test data when true labels are available.

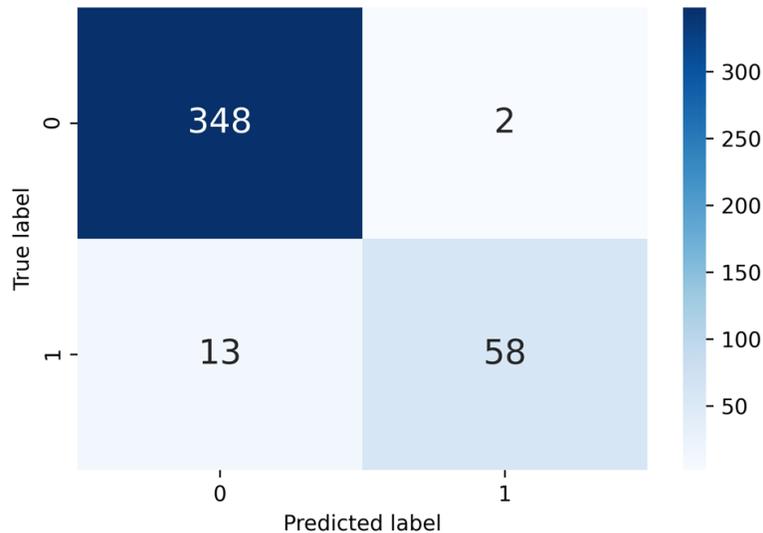

**Figure 11:** Confusion matrix of classification

The CNN correctly classified 348 healthy audio samples and 58 COPD samples, resulting in 406 accurate predictions. However, the model misclassified 2 healthy samples as COPD and produced 15 total errors.

*3.3 Comparison of result*

In this work, two methodologies were applied to evaluate the proposed CNN model. Table 1 presents studies that employed different CNN architectures, which are compared with the proposed model. The proposed model achieved 97% accuracy without 10-fold cross-validation, demonstrating a clear improvement over existing approach.

**Table 1**: Model's result comparison

| Reference | Algorithm | Accuracy (%) |
|---|---|---|
| **This paper** | **Custom Convolutional Neural Network** | **96.0** |
| Ref [17] | VGG-19 | 89.5 |
| Ref [18] | CNN | 93.4 |
| Ref [19] | YOLO | 88.0 |

This study achieved 96% accuracy with a custom CNN model, whereas Ref. [18] reported 93% accuracy using a CNN. The YOLO models in Refs. [17] and [19], which incorporated VGG-19, obtained lower accuracy than the proposed model.

**4 Conclusion**

In this paper, we propose an efficient CNN-based assistive model to support clinical experts in detecting COPD from respiratory sounds. For feature extraction, we employed the Librosa library using MFCC, Mel-Spectrogram, Chroma, Chroma (ConstantQ), and Chroma CENS. Based on systematic evaluation on the dataset, MFCC achieved superior accuracy in distinguishing COPD from non-COPD compared to the other features.

For future work, the system can be extended to identify additional conditions such as cardiovascular failure from heartbeat sounds and asthma from lung sounds. The current framework can also be enhanced to assess disease severity. Furthermore, data augmentation strategies may improve model performance. Integration with a respiratory monitoring system [31] could further simplify real-time COPD detection. Ensuring robustness against adversarial attacks and strengthening privacy protection [22] will also be essential. While the current system demonstrates high reliability in differentiating COPD from non-COPD, deployment across vehicles and other real-world environments will enable continuous monitoring of pulmonary conditions in real time.

**Acknowledgement:** Headings must be left-justified, bold, unnumbered, and capitalized with only the first letter of each word. The text following these headings should follow the main manuscript formatting.

In the **Acknowledgement** section, authors may thank individuals or institutions that contributed to the work but must not include themselves.

*Funding Statement*: Authors should describe sources of funding that have supported their work, including specific grant numbers, initials of authors who received the grant, and the URLs to sponsors' websites. If there is no funding support, please write "The author(s) received no specific funding for this study."

**Conflicts of Interest:** Authors must declare all and any conflicts of interest. If there are noconflicts of interest, it should also be declared as such: "The authors declare that they have no conflicts of interest to report regarding the present study."

**References**

[1] J Ahmed, S.Vesal, F.Durlak, R. Kaergel, N. Ravikumar, M. Rémy-Jardin, A. Maier, T. Tolxdorff, T. Deserno, H. Handels, A. Maier. 2020. COPD classification in CT images using a 3D convolutional neural network. In: Maier-Hein K, Palm C, eds. Bildverarbeitung für Die Medizin 2020—Informatik Aktuell. Wiesbaden: Springer Vieweg, 39–45.

[2] G. Altan, Y. Kutlu, N. Allahwardi. 2019. Deep learning on computerized analysis of chronic obstructive pulmonary disease. IEEE Journal of Biomedical and Health Informatics 24(5):1344–1350 DOI 10.1109/JBHI.2019.2931395.

[3] G. Altan, Y. Kutlu, A.O. Pekmezci, S. Nural. 2018. Deep learning with 3D-second order difference plot on respiratory sounds. Biomedical Signal Processing and Control 45:58–69 DOI 10.1016/j.bspc.2018.05.014.

[4] J.L. Amaral, A.J. Lopes, J.M. Jansen, A.C. Faria, P.L. Melo. 2012. Machine learning algorithms and forced oscillation measurements applied to the automatic identification of chronic obstructive pulmonary disease. Computer Methods and Programs in Biomedicine 105(3):183–193 DOI 10.1016/j.cmpb.2011.09.009.

[5] M. Asaithambi, S.C. Manoharan, S. Subramanian. 2012. Classification of respiratory abnormalities using adaptive neuro-fuzzy inference system. In: Pan J-S, Chen S-M, Nguyen NT, eds. Intelligent Information and Database Systems. Berlin: Springer, 65–73.

[6] M.Aykanat, O. Kılıç, B. Kurt, S. Saryal. 2017. Classification of lung sounds using convolutional neural networks. EURASIP Journal on Image and Video Processing 2017(1):65 DOI 10.1186/s13640-017-0213-2.

[7] A. Badnjevic, L. Gurbeta, E. Custovic. 2018. An expert diagnostic system to automatically identify asthma and chronic obstructive pulmonary disease in clinical settings. Scientific Reports 8(1):11645 DOI 10.1038/s41598-018-30116-2.

[8] Z. Bai, Y. Li, M. Woźniak, M. Zhou, D. Li. 2020. DecomVQANet: decomposing visual question answering deep network via tensor decomposition and regression. Pattern Recognition 110:107538 DOI 10.1016/j.patcog.2020.107538.

[9] C.C. Bellos, A. Papadopoulos, R. Rosso, D.I.Fotiadis. 2014. Identification of COPD patients' health status using an intelligent system in the CHRONIOUS wearable platform. IEEE Journal of Biomedical and Health Informatics 18(3):731–738 DOI 10.1109/JBHI.2013.2293172.

[10] D. Chamberlain, R. Kodgule, D. Ganelin,V. Miglani, R.R. Fletcher. 2016. Application of semi-supervised deep learning to lung sound analysis. In: 38th Annual International

Conference of the IEEE Engineering in Medicine and Biology Society (EMBC). Piscataway: IEEE, 804–807.

[11] R. Du, S. Qi, J. Feng, S. Xia, Y. Kang, W. Qian, Y.D. Yao. 2020. Identification of COPD from multi-view snapshots of 3D lung airway tree via deep CNN. IEEE Access 8:38907–38919 DOI 10.1109/ACCESS.2020.2974617.

[12] O.Er, F.Temurtas. 2008. A study on chronic obstructive pulmonary disease diagnosis using multilayer neural networks. Journal of Medical Systems 32(5):429–432 DOI 10.1007/s10916-008-9148-6.

[13] M.A. Fernandez-Granero, D. Sanchez-Morillo, A. Leon-Jimenez. 2018. An artificial intelligence approach to early predict symptom-based exacerbations of COPD. Biotechnology & Biotechnological Equipment 32(3):778–784 DOI 10.1080/13102818.2018.1437568.

[14] K.L. Khatri, L.S.Tamil. 2018. Early detection of peak demand days of chronic respiratory diseases emergency department visits using artificial neural networks. IEEE Journal of Biomedical and Health Informatics 22(1):285–290 DOI 10.1109/JBHI.2017.2698418.

[15] ICBHI. 2017. ICBHI challenge. Available at https://bhichallenge.med.auth.gr/ICBHI_2017_ Challenge (accessed 21 May 2020).

[16] K.L. Khatri, L.S. Tamil. 2018. Early detection of peak demand days of chronic respiratory diseases emergency department visits using artificial neural networks. IEEE Journal of Biomedical and Health Informatics 22(1):285–290 DOI 10.1109/JBHI.2017.2698418.

[17] Mahmud, M. Z., Reza, M. S., Alve, S. R., Islam, S., & Fahmid, N. (2024). Advance transfer learning approach for identification of multiclass skin disease with LIME explainable AI technique. medRxiv. https://doi.org/10.1101/2024.12.02.24318311

[18] Islam, S., Mahmud, M. Z., Alve, S. R., & Chowdhury, M. M. U. (2024). Deep learning approach for enhancing oral squamous cell carcinoma with LIME explainable AI technique. arXiv preprint arXiv:2411.14184. https://arxiv.org/abs/2411.14184

[19] Alve, S. R. (2024). Deep learning and hybrid approaches for dynamic scene analysis, object detection, and motion tracking. arXiv preprint, arXiv:2412.05331.

[20] Y. Liu, Y. Lin, S. Gao, H. Zhang, Z. Wang, Y. Gao, G. Chen. 2017. Respiratory sounds feature learning with deep convolutional neural networks. In: 2017 IEEE 15th International Conference on Dependable, Autonomic and Secure Computing, 15th International Conference on Pervasive Intelligence and Computing, 3rd International Conference on Big Data Intelligence and Computing and Cyber Science and Technology Congress (DASC/PiCom/DataCom/CyberSciTech). Piscataway: IEEE, 170–177.

[21] Mahmud, M. Z., Islam, S., Alve, S. R., & Pial, A. J. (2024). Optimized IoT intrusion detection using machine learning technique. arXiv preprint arXiv:2412.02845

[22] S. Patil, J. Shashank, P. Deepali. 2020. Enhanced privacy preservation using anonymization in IoT-enabled smart homes. In: Smart Intelligent Computing and Applications. Singapore: Springer, 439–454.

[23] D. Perna. 2018. Convolutional neural networks learning from respiratory data. In: 2018 IEEE International Conference on Bioinformatics and Biomedicine (BIBM). Madrid: IEEE, 2109–2113.

[24] D. Perna, A. Tagarelli. 2019. Deep auscultation: predicting respiratory anomalies and diseases via recurrent neural networks. arXiv. Available at http://arxiv.org/abs/1907.05708.

[25] L. Pham, I. McLoughlin, H. Phan, M. Tran, T. Nguyen, R. Palaniappan. 2020. Robust deep learning framework for predicting respiratory anomalies and diseases. arXiv. Available at http://arxiv.org/ abs/2002.03894.

[26] A.V. Radogna, N.Fiore, M.R. Tumolo, V. De Luca, L.T. De Paolis, R. Guarino, C.G.Leo, P. Mincarone, E. Sabato, F. Satriano, S. Capone, S. Sabina. 2019. Exhaled breath monitoring during home ventilo-therapy in COPD patients by a new distributed tele-medicine system. Journal of Ambient Intelligence and Humanized Computing 5:1–9.

[27] B.M. Rocha, D. Filos, L. Mendes, G. Serbes, S. Ulukaya, Y.P. Kahya, N. Jakovljevic, T.L.Turukalo, I.M.Vogiatzis, E. Perantoni, E. Kaimakamis, P. Natsiavas, A. Oliveira, C. Jácome, A. Marques, N. Maglaveras, R. Pedro Paiva, I. Chouvarda, P. De Carvalho. 2019. An open access database for the evaluation of respiratory sound classification algorithms. Physiological Measurement 40(3):035001.

[28] A. Shaikh, S. Patil. 2018. A survey on privacy enhanced role based data aggregation via differential privacy. In: 2018 International Conference On Advances in Communication and Computing Technology. Piscataway: IEEE, 285–290.

[29] J. Weese, C. Lorenz. 2016. Four challenges in medical image analysis from an industrial perspective. Medical Image Analysis 33:44–49 DOI 10.1016/j.media.2016.06.023.

[30] R. Wierenga. 2020. An empirical comparison of optimizers for machine learning models. Available at https://heartbeat.fritz.ai/an-empirical-comparison-of-optimizers-for-machinelearning-models-b86f29957050.

[31] S. Manoharan, M. Veezhinathan, S. Ramakrishnan. 2008. Comparison of two ANN methods for classification of spirometer data. Measurement Science Review 8(3):535 DOI 10.2478/v10048-008-0014-y

[32] Mahmud, M. Z., Alve, S. R., Islam, S., & Khan, M. M. (2024). SDN intrusion detection using machine learning methods. arXiv preprint, arXiv:2411.05888